\documentclass[fleqn,usenatbib]{mnras}

\usepackage{newtxtext,newtxmath}
\usepackage[T1]{fontenc}
\usepackage{ae,aecompl}

\usepackage{graphicx}	% Including figure files
\usepackage{amsmath}	% Advanced maths commands
\usepackage{amssymb}	% Extra maths symbols

\usepackage{natbib}
\usepackage{bm}
\usepackage[dvipdfmx]{color}

\usepackage{hyperref}
\usepackage{color}
\usepackage{ulem}
\usepackage{comment}
\usepackage{braket}
\newcommand{\HI}{\textrm{H}_{\textrm{I}}}
\newcommand{\Tb}{\delta T_{\textrm{b}}}

\title[%Short title, max. 45 characters
	AP test with 21cm Void]{The Alcock Paczynski test with 21cm intensity field}

\author[Endo, Tashiro \& Nishizawa]
{Takao Endo$^{1}$\thanks{Email: endou.takao@a.mbox.nagoya-u.ac.jp},
Hiroyuki Tashiro$^{1}$\thanks{Email: hiroyuki.tashiro@nagoya-u.jp}
and Atsushi J. Nishizawa$^{1,2}$\thanks{Email: atsushi.nishizawa@iar.nagoya-u.ac.jp}\\
$^{1}$ Graduate School of Science, Nagoya University, Aichi 464-8602, Japan,\\
$^{2}$ Institute for Advanced Research, Nagoya University, Aichi 464-8602, Japan,\\
}

\date{Accepted XXX. Received YYY; in original form ZZZ}

\pubyear{2015}

\begin{document}
\label{firstpage}
\pagerange{\pageref{firstpage}--\pageref{lastpage}}
\maketitle

% Abstract of the paper
\begin{abstract}
Feasibility of
the Alcock Paczynski (AP) test 
by stacking voids in the 21cm line intensity field is presented.
We analyze the Illstris-TNG simulation to obtain the 21cm signal map.
We then randomly distribute particles depending on the 21cm intensity field 
to find voids by using publicly available code, \texttt{VIDE}.
As in the galaxy clustering, the shape of the stacked void in the 21cm field is squashed
along the line of sight due to the peculiar velocities
in redshift-space, although it becomes spherical in real-space. 
The redshift-space distortion for the stacked void weakly depends on redshift and we show that the dependency can be well described by the linear prediction, with the amplitude of the offset being free parameters.
We find that the AP test using the stacked voids in a 21cm intensity map is feasible and the parameter estimation on $\Omega_{\rm m}$ and $w$ is unbiased.
%This is a simple template for authors to write new MNRAS papers.
%The abstract should briefly describe the aims, methods, and main results of the paper.
%It should be a single paragraph, not more than 250 words (200 words for Letters).
%No references should appear in the abstract.
\end{abstract}

% Select between one and six entries from the list of approved keywords.
% Don't make up new ones.

%%%%%%%%%%%%%%%%%%%%%%%%%%%%%%%%%%%%%%%%%%%%%
%%%%%%%%%%%%%%%%% BODY OF PAPER %%%%%%%%%%%%%%%%%%
%%%%%%%%%%%%%%%%%%%%%%%%%%%%%%%%%%%%%%%%%%%%%

%%%%%%%%%%%%%%%%%%%%%%%%%%%%%%%%%%%%%%%%%%%%%
%%%%%%%%%%%%%%%%% START SECTIONS %%%%%%%%%%%%%%%%%%
%%%%%%%%%%%%%%%%%%%%%%%%%%%%%%%%%%%%%%%%%%%%%
\section{Introduction}
\label{sec:introduction}
Recent developments in observations drive the remarkable progress in cosmology. 
Based on the fruitful observational data, we can obtain the precise constraint on the theoretical scenarios and models of the Universe.
On the other hand, these observations throw up new challenges to us for understanding the Universe. 
One of them is the discovery of the accelerating expansion of the Universe~\citep{Riess:1998, Perlmutter:1999}
The most accepted hypothesis to explain this acceleration is 
dark energy. Dark energy has large negative pressure 
and can generate a repulsive gravitational force to accelerate the
cosmic expansion~\citep{Weinberg:2013}.
The joint analysis among the CMB anisotropy measurement and galaxy
surveys favor the existence of dark energy~\citep{Planck2018:cosmology}.
However, there are not sufficient observations to
determine its nature precisely.
Therefore it is required to make further efforts to perform
CMB anisotropy observations and galaxy surveys more precisely and conduct cosmological observations independent of them.

The Alcock Paczynski (AP) test~\citep{Alcock:1979} is
also, one of the independent observations to probe the expansion history
of the Universe.  
For the success of the AP test, it is required to know the physical size
of observation objects along the line of sight
and angular direction, or the ratio between them.
Therefore, the objects whose shapes are spherical or isotropic are the ideal
targets for the AP test.
There are many works to study the AP test with cosmological objects.
The peak scale of the Baryon Acoustic Oscillation \citep{HS:1996, Eisenstein:2005},
might be one of the best candidates for the AP test 
because the peak scale is isotropic and well determined by the linear
theory and observations~\citep{Komatsu:2011,Planck2018:cosmology}.
Since objects whose shapes are statistically isotropic are also preferable for the AP test,
the correlation function of galaxies and the matter power spectrum
are studied as objects for the AP test \citep{Matsubara:1996,Ballinger:1996,Hu:2003,Seo:2003,Matsubara:2004,Glazbrook:2005}.

Now cosmic voids have been paid attention as cosmological structures to probe the Universe.
While some authors investigated the relation of
void size distribution to the cosmological
models~\citep{Clampitt:2013,Pisani:2015,Zivick:2016,Endo:2018,Verza:2019}.
the application of voids also has been studied.
The AP test with voids was originally proposed by~\cite{Ryden:1995}.
\cite{Ryden:1995} demonstrated the AP test with voids
in the toy model where individual shapes of voids are assumed to be spherical. 
However, their real shapes are far from spherical both in matter density simulations \citep{Platen:2008}
and galaxy surveys \citep{Nadathur:2016}.  
To solve this shape problem,
\cite{Lavaux:2012} has proposed to use
the statistical shapes of voids in the AP test.
They have argued that the stacked voids can be applied to the AP test
because they are expected to be spherical based on the cosmological principle, the statistical isotropy of the Universe.

Up to the present, there have been several studies on this topic.
\cite{Sutter:2012} first applied the stacking method to the AP test
for the constraint on the matter-energy density parameter~$\Omega_{\rm m}$
with voids detected by the galaxy survey, the Sloan Digital Sky Survey Data Release (SDSS) Data Release 7~\citep{SDSS:DR7}.
However, their AP test could not provide the constraint on $\Omega_{\rm m}$,
because of the insufficient number of observed voids at that time.
Later, \citep{Sutter:2014} and \cite{Mao:2017}
revisited the AP test with stacked voids detected from SDSS Data Release
10~\citep{SDSS:DR10} and 12~\citep{SDSS:DR12}.
Although they have obtained the constraint on $\Omega_{\rm m}$,
the constraints are weak compared with those from other cosmological observations.

The precision and tightness of the constraint by the AP test depends
on the statistical spherical symmetry of void shapes in an observation data set.
To recover the spherical symmetry, a large number of void samples are required.
For this reason, we need a huge survey volume, covering a wide range of the sky and
reaching deep redshift.
The future 21cm intensity mapping survey by the Square Kilometre Array~(SKA)
is planed to investigate the huge volume overwhelming the conventional galaxy
surveys~\citep{Santos:2015,SKA:redbook2018}.
In the previous works, voids are found 
in the distribution maps of galaxies produced by galaxy surveys.
On the other hand,
SKA traces the distribution of neutral hydrogen $\HI$ in large scale
structure by measuring the 21cm line signals caused by their hyperfine
structure transition.
The void regions also could be observed as low-intensity regions
in the 21cm intensity maps.
If we detect the voids from the $\HI$ intensity fields, we will obtain the number of voids more than the galaxy survey.  

In this paper, we propose the 
AP test with void regions in the 21cm intensity map.
We use the contour of the intensity
to find out voids in the 21cm intensity map.
When we plot the contour for the appropriate intensity,
we can find that the sizable low-density regions are enclosed by the contour.
Similar to voids in the galaxy survey map,
we can assume that voids in the 21cm intensity map
are statistically spherical according to the cosmological principle.
Using the state-of-the-art cosmological hydrodynamics simulation data,
we conduct the void findings by the contour in the 21cm intensity
map and show that our assumption about the statistical spherical symmetry is valid.
We also perform the AP test with stacking the found voids
and demonstrate that the AP test of the stacked voids in the intensity maps 
can provide the tight constraint with the survey volume planned in the
SKA project.

%%%%%%%%%%%%%%%%%%%%%%%%%%%%%%%%%%%%%%%%%%%%%%%%%%
%%%%%%%%%%%%%%%%%%%%%%%%%%%%%%%%%%%%%%%%%%%%%%%%%%
\section{Method}
\label{sec:method}
%%%%%%%%%%%%%%%%%%%%%%%%%%%%%%%%%%%%%%%%%%%%%%%%%
The aim of this paper is to demonstrate the feasibility of the AP test
using the contour lines on the 21cm intensity map.
In this section,
we present how we construct the 21cm intensity map from
numerical simulation data.

\subsection{Hydrodynamical simulation}
\label{ssec:Data} % used for referring to this section from elsewhere
%%%%%%%%%%%%%%%%%%%%%%%%%%%%%%%%%%%%%%%%%%%%%%%%%%
The 21cm signal depends on the number density and temperature of the intergalactic medium (IGM) gas. 
Therefore, to predict the cosmological 21cm intensity map properly,
the cosmological hydrodynamics simulation is required.
In this paper, we used the data of the state-of-the-art cosmological magnetohydrodynamical simulation,
called IllustrisTNG~\citep{Pillepich:2018, Naiman:2018,Springel:2018,Nelson:2018,Mariacci:2018,Nelson:2019}.
The simulation solves the gravitational force by a Tree-PM
scheme~\citep{Xu:1995} and
hydrodynamics by running a moving-mesh code,
the AREPO code~\citep{Springer:2010}.
The simulation also includes a wide range of astrophysical processes,
which provide large impacts on the gas state in the IGM, such as
radiative cooling, star formation, stellar evolutions, stellar feedback,
growth of supermassive black holes and their feedbacks and magnetic fields~\citep{Nelson:2019}.
Among the series of the Illustris simulation results,
we adopt the IllustrisTNG300-3 data.
The volume of the simulation box is $(205 {\rm cMpc}/h)^3$
and the dynamics of $2\times625^3$ dark matter particles and gas cells were solved simultaneously.

Using this data, we construct the 21cm intensity maps.  
The 21cm intensity depends on the physical state of the IGM gas.
To evaluate the 21cm signal simply,
we prepare a $256^3$ grid space in the simulation box and calculate the physical gas quantities in each cell with the
Cloud-in-Cell scheme. 
The simulation treats the gas component as a finite size cell but
 the size of the cell is enough smaller than the grid separation.
Therefore, we can consider the gas as a particle and neglect the finite size effect of the gas.
In the simulation, the cosmological parameters are based on
the Planck 2015 measurements~\citep{Planck2015:cosmology}:
$\Omega_{\textrm{m}}=0.3089$, $\Omega_{\Lambda}=0.6911$,
$\Omega_{\textrm{b}} = 0.0486$ , $\sigma_8=0.8159$, $n_s=0.9667$
and $H_0 = 100h\textrm{km}\ \textrm{s}^{-1}\ \textrm{Mpc}^{-1}$ with $h=0.6774$.

\begin{figure*}
 	\includegraphics[width=0.45\linewidth]{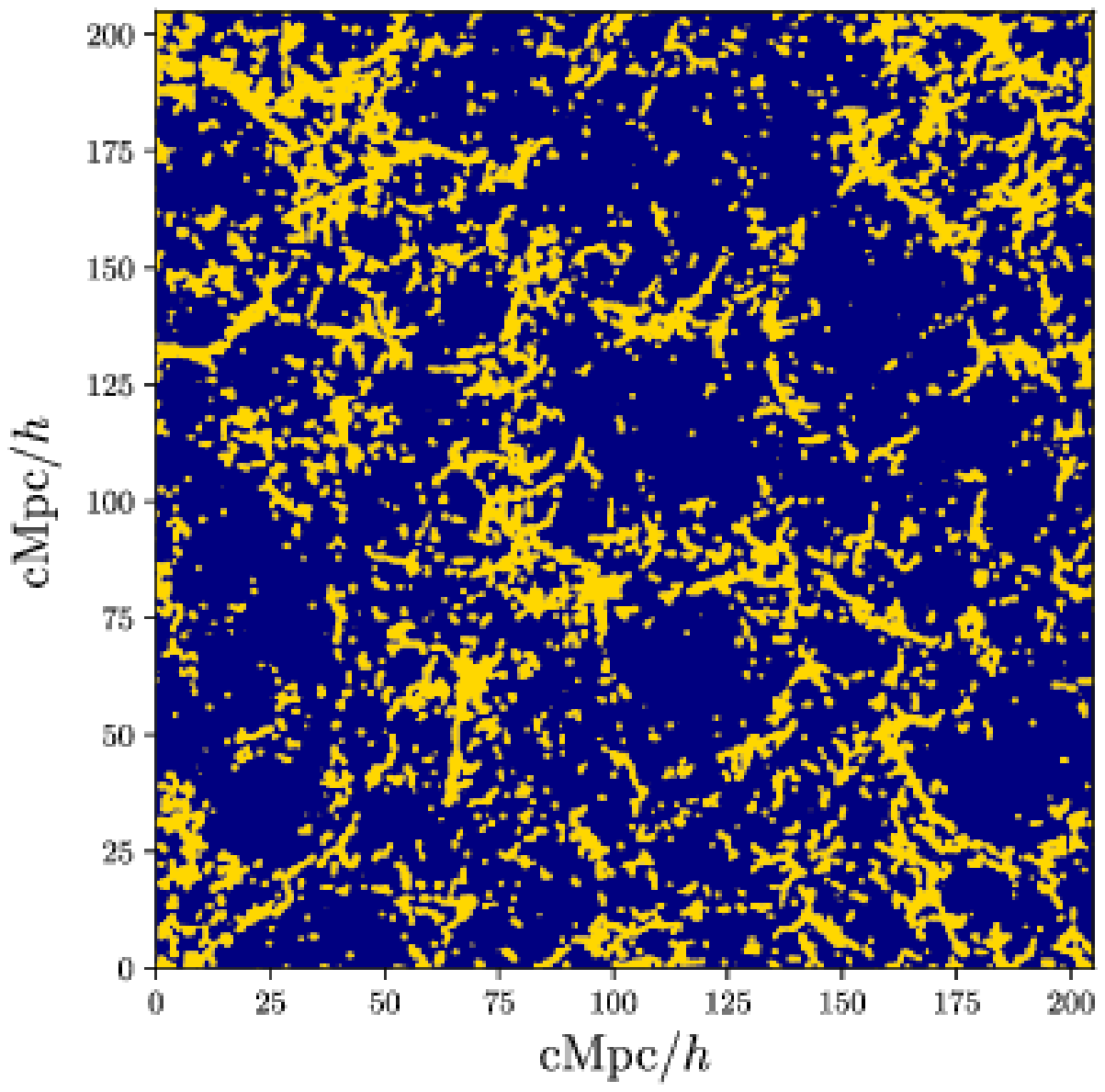}
 	\includegraphics[width=0.45\linewidth]{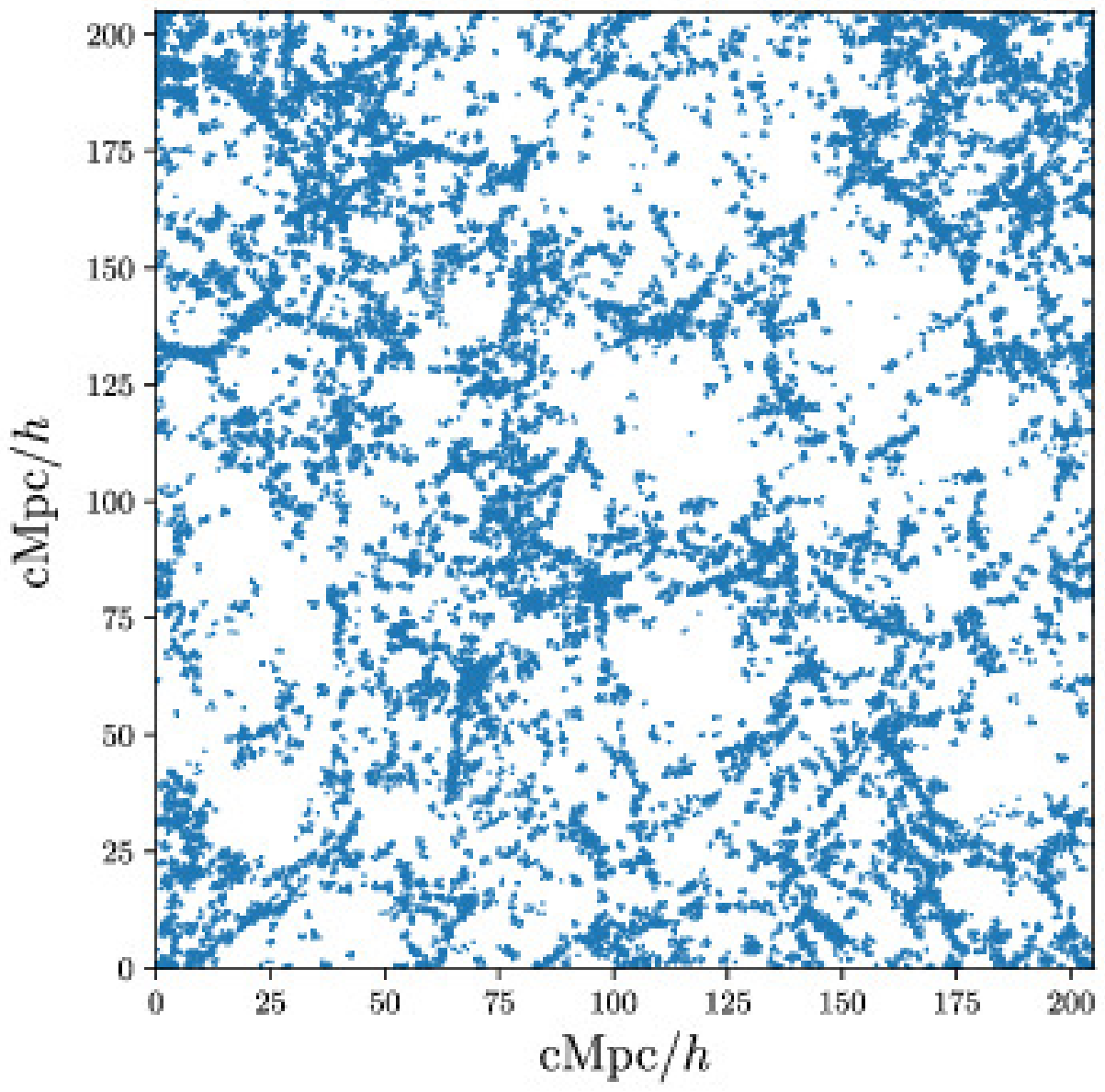}
	 \caption{The intensity contour map made from the IllustrisTNG simulation (left panel)
  and the mock $\HI$ particle distribution (right panel) made from the left panel by means of the method
  described in section \ref{sec:voidFinding}. 
  The contour level of the left panel is average brightness temperature $\braket{\Tb}$ so that we can regard as the dark regions as voids while bright regions as filaments or halos.  
  In the right panel, we can see that the distribution of mock particles traces the shape of void regions in the left panel.
 \label{fig:TbMap}}
\end{figure*}

%%%%%%%%%%%%%%%%%%%%%%%%%%%%%%%%%%%%%%%%%%%%%%%%%%
\subsection{$\textrm{H}_{\textrm{I}}$ intensity map}
\label{ssec:HIintensityMap} % used for referring to this section from elsewhere
%%%%%%%%%%%%%%%%%%%%%%%%%%%%%%%%%%%%%%%%%%%%%%%%%%

%%%%%%%%%%%%%%%%%%%%%%%%%%%%%%%%%%%%%%%%%%%%%%%%%
\subsubsection{The Differential Brightness Temperature}
\label{ssec:deltaTb} % used for referring to this section from elsewhere
%%%%%%%%%%%%%%%%%%%%%%%%%%%%%%%%%%%%%%%%%%%%%%%%%%
Neutral hydrogen atom emits or absorbs specific electromagnetic radiation whose wavelength
corresponds to 21cm when its spin configuration between proton and
electron, i.e. hyperfine structure, changes.
When we observe the cosmological 21cm signal,
we measure it as a difference from
the CMB brightness temperature~\citep{Furlanetto:2006},
\begin{align}
	\delta T_{\rm b} = \frac{ \left[ T_{\rm s}(z) - T_{\rm CMB} (z) \right] \left( 1- e^{-\tau_{10}} \right)}{1+z},
\end{align}
where $\delta T_{\rm b}, T_{\rm CMB}$ and $T_{\rm s}$ are the differential brightness temperature,
the CMB temperature and the spin temperature which is defined as the
population ratio between the upper and lower states of the hyperfine structure in neutral hydrogen, respectively.
The optical depth $\tau_{10}$ for the 21cm line is
\begin{align}
	\tau_{10} 
	=
	\int ds \frac{3}{32 \pi}
	\frac{h_{\rm p} c^2 A_{10}}{\nu_{*} k_{\rm B}}
	\frac{n_{\rm H_I}}{T_{\rm s}} \phi(\nu),
\label{eq:opt_21}
\end{align}
where $h_{\rm p}$ is the Planck constant, $c$ is the speed of light,
the subscript~$10$ represents the value corresponding to the
energy gap between hyperfine structure levels,~$\nu_{*}=1.4~{\rm GHz}$, $A_{10}=2.85\times10^{-15}~{\rm
s}^{-1}$ is the Einstein A-coefficient of the transition, $k_{\rm B}$ is the Boltzmann constant,
and $n_{\HI}$ is the number density of neutral hydrogen.
In equation~\eqref{eq:opt_21},
$\phi(\nu)$ is a line profile which has a peak at the frequency~$\nu_{*}$ and satisfies the normalization,~$\int \phi(\nu) d\nu = 1$.
The line profile suffers the broadening due to the velocity of the $\HI$ gas in the line-of-sight direction, $V_{\parallel}$,
which includes the velocities due to the Hubble expansion, the peculiar velocity and the thermal velocity of the gas.
Accordingly, the width of the profile is broadened and the amplitude is suppressed by ${c}/{ \nu_{*} V_{\parallel}}$ around the peak frequency.
Performing the integration in equation~\eqref{eq:opt_21} with this broadening, we obtain
\begin{align}
	\Tb = \frac{3c^3}{32 \pi} \frac{A_{10} h_{\textrm{p} } n_{\HI}}{k_{\textrm{B}} \nu_{*}^2} \frac{1}{(1+z)^2 |dV_{\parallel}/dx_{\parallel}| } \left( 1 - \frac{T_{\textrm{CMB}}(z)}{T_{\textrm{s}}(z)} \right),
	\label{eq:delta_Tb}
\end{align}
where $dV_{\parallel}/dx_{\parallel}$ is the gas velocity gradient along the line-of-sight.
In the redshifts and scales that we are interested in throughout this paper,
we assume that the dominant contribution to $V_{\parallel}$ comes from the Hubble expansion~\citep{Horii:2017}.
Thus we can approximate the velocity gradient as
\begin{align}
\frac{dV_{\parallel}}{ dx_{\parallel}} \approx \frac{H(z)}{(1+z)}. 
\end{align}

%%%%%%%%%%%%%%%%%%%%%%%%%%%%%%%%%%%%%%%%%%%%%%%%%%
\subsubsection{Spin temperature}
\label{ssec:Tspin} % used for referring to this section from elsewhere
%%%%%%%%%%%%%%%%%%%%%%%%%%%%%%%%%%%%%%%%%%%%%%%%%%
The spin temperature is determined by the balance related to the spin flipping physics in the hyperfine structure including the coupling with CMB photons, thermal collisions with electron and other atoms and the pumping by the background Ly-$\alpha$ photons, which is known as the Wouthuysen-Field effect~\citep{Wouthuysen:1952, Field:1958}.
As a result, the spin temperature can be expressed as~\citep{Field:1959, Furlanetto:2006}
\begin{align}
T_{\rm s}^{-1} = \frac{T^{-1}_{\rm CMB} + x_{\rm c} T^{-1}_{\rm K} + x_{\alpha} T_{\alpha}^{-1} }{ 1 + x_{\rm c} + x_{\alpha}},
\end{align}
where $T_{\textrm{K}}$ and $T_{\alpha}$ are the kinetic temperature of $\textrm{H}_{\textrm{I}}$ gas and the color-temperature of the Ly-$\alpha$ radiation field.
Throughout this work, we assume $T_{\alpha} \simeq T_{\textrm{K}}$.
In the equation, $x_{\rm c}$ and $x_{\alpha}$ are the coupling coefficients for the thermal collisions and the Ly-$\alpha$ photons respectively.
The coupling coefficient $x_{\rm c}$ is described by
\begin{align}
	x_{\rm c} = \frac{T_{*}}{A_{10} T_\textrm{CMB}} \left( n_{\textrm{H}_{\textrm{I}}} \kappa_{\textrm{H}_{\textrm{I}}} + n_{\textrm{p}}\kappa_{\textrm{p}} + n_{\textrm{e}} \kappa_{\textrm{e}} \right)
\end{align}
where $T_{*} = h_{\textrm{p}} \nu_{*}/k_{\textrm{B}} = 0.068$ K, and $n_{\textrm{p}}$ and $n_{\textrm{e}}$ are the number densities of protons and electrons respectively.
We need collision rates $\kappa_{\textrm{H}_{\textrm{I}}}$, $\kappa_{\textrm{p}} $ and $\kappa_{\textrm{e}}$ to calculate $x_{\alpha}$.  
We follow the same manner in \cite{Kuhlen:2006} which uses a fitting formulae based on \cite{Field:1958}, \cite{smith:1966}, \cite{Allison:1969} ,\cite{Liszt:2001} and \cite{Zygelman:2005}.
In the unit of [cm$^3$/s], these coefficients are given in

\begin{align}
	&\kappa_{\HI} = 3.1\times 10^{-11} T_{\rm K}^{0.357} \exp \left( -\frac{-32}{T_{\rm K}} \right),\\
	&\kappa_{e} = 
	\begin{cases}
  10^{-9.607 + 0.5 \log(T_K) \exp\left( -(\log T_{K\rm })^{4.5}/1800 \right)} \ \ &(T_{\rm K} \leq 10^4), \\
  \kappa_e(T_{\rm K}=10^4) &(T_{\rm K} > 10^4),
	\end{cases}\\
	&\kappa_{p} = 3.2 \kappa_H,
\end{align}
where $T_{\rm K}$ is in the unit of [K].

The coupling coefficient $x_{\alpha}$ can be approximated as \citep{Furlanetto:2006}
\begin{align}
	x_{\alpha} = S_{\alpha} \frac{J_{\alpha} }{J^{\textrm{c} }_{\nu} },
\end{align}
where $J^{\textrm{c} }_{\nu} = 1.165 \times 10^{-10} [(1+z)/20]$ $\textrm{cm} ^2 \textrm{s}^{-1} \textrm{Hz}^{-1} \textrm{sr}^{-1}$,
and $J_{\alpha}$ is the background Ly-$\alpha$ photon intensity. The scattering amplitude factor, $S_\alpha$, is approximated as
\begin{align}
	S_{\alpha} \sim \exp \left[ -0.803 T_{\rm K} ^{-2/3} \left( \frac{10^{-6}}{ \gamma} \right)^{1/3} \right],
\end{align}
where
\begin{align}
	\gamma = \frac{H(z) \nu_{\alpha}}{\chi _{\alpha} n_{\textrm{H}_{\textrm{I} } } c},
\end{align}
with the Lyman-$\alpha$ frequency $\nu_{\alpha}=2.47 \times 10^{15} \textrm{Hz}$ and
\begin{align}
	\chi_{\alpha} = \frac{\pi e^2}{m_{\textrm{e}} c} f_{\alpha},
\end{align}
where $e$ is the electron charge, $m_{\textrm{e}}$ is the electron mass and $f_{\alpha} = 0.4162$ is the oscillator length.
For $J_{\alpha}$, we adopt the results in~\cite{Haardt:2012}.

%%%%%%%%%%%%%%%%%%%%%%%%%%%%%%%%%%%%%%%%%%%%%%%%%%
\subsubsection{Intensity contour map from the Illustris data}
\label{ssec:intensity} % used for referring to this section from elsewhere
%%%%%%%%%%%%%%%%%%%%%%%%%%%%%%%%%%%%%%%%%%%%%%%%%%
Following section~\ref{ssec:deltaTb} and \ref{ssec:Tspin},
we obtain the 21cm intensity~(the differential brightness temperature) maps from the IllustrisTNG300-3 simulation data.
In the left panel of Figure~\ref{fig:TbMap}
we show the contour map of the intensity at $z=1$.
Here we take one contour line corresponding to the averaged intensity $\braket{\Tb}$ in the simulation box
to make the contour map.
The bright color regions have larger intensity than the average value,
while the dark color regions have lower intensity.
Since the intensity is proportional to the $\HI$ density, the dark regions can be identified as voids.

Each dark region surrounded by the bright filamentary structures
is not spherical. However, the cosmological principle allows us to make a hypothesis that the dark regions are statistically spherical
and to apply the stacked shape of the low-intensity regions to the Alcock-Paczynski test.
To extract each shape of the dark region as a void in the contour map, we adopt publicly available code, \texttt{VIDE}~\citep{Sutter:vide}.  
In the next subsection, we describe how we apply the void finding to our contour maps.
%%%%%%%%%%%%%%%%%%%%%%%%%%%%%%%%%%%%%%%%%%%%%%%%%%

%%%%%%%%%%%%%%%%%%%%%%%%%%%%%%%%%%%%%%%%%%%%%%%%%%
\subsection{Void finding}
\label{sec:voidFinding} % used for referring to this section from elsewhere
%%%%%%%%%%%%%%%%%%%%%%%%%%%%%%%%%%%%%%%%%%%%%%%%%%
To trace the shape of the dark~(low-intensity) regions in the left panels of Figure~\ref{fig:TbMap},
we exploit the \texttt{VIDE} algorithm~\citep{Sutter:vide}.
This algorithm is an updated version of the \texttt{ZOBOV} algorithm~\citep{Neyrinck:zobov}
which uses the Voronoi tessellation to make a density field and
applies the watershed method~\citep{Platen:2007} to detect void regions.

The \texttt{VIDE} algorithm finds out voids based on a particle distribution.
Therefore, we need to construct the particle distribution from the
21cm intensity contour map as shown in Figure~\ref{fig:TbMap}. 
Since the dark regions are encircled by the bright regions,
we put a particle in each grid of the bright regions.
In the right panel of Figure~\ref{fig:TbMap},
we show the mock particle distribution obtained from the left panel.
One can see that the mock particle distribution traces the outlines of the dark regions on the left panel.

\begin{figure}
 \includegraphics[width=\linewidth]{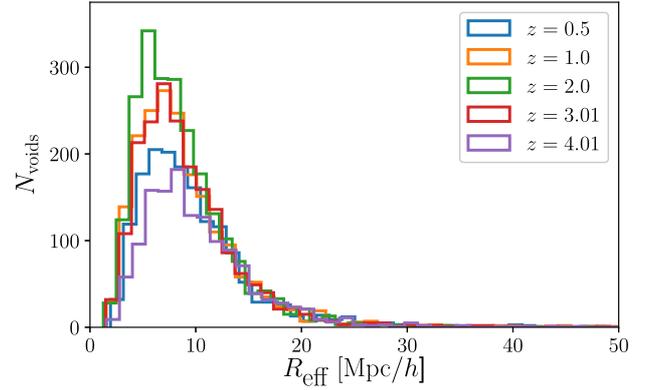}
 	\caption{
   The size distribution of the voids at different redshifts. In our simulation, voids are populated at $5 < R_{\rm eff} < 10 {\rm Mpc}/h$ and increase till $z=2$ and decrease afterword.
 	\label{fig:numberFunction}}
\end{figure}

Now we can obtain the statistical property of the dark regions as ``voids'' 
using the \texttt{VIDE} algorithm with the mock particle distribution.
We show the obtained void abundance in Figure~\ref{fig:numberFunction} where
the $x$-axis represents the effective radius of a void as
\begin{align}
	R_{\textrm{eff}}= \left( \frac{3}{4\pi}V \right)^{1/3},
	\label{eq:rEff}
\end{align}
where $V$ is the volume of the void region evaluated by the \texttt{VIDE} algorithm.
To represent the redshift dependence, we also provide the void abundance at different redshifts, $z=0.5,~1,~2,~3$ and $4$
by using the snapshot data at those redshifts.

The abundances of voids in our catalogs have peaks between $R_{\textrm{eff}} = 5 \textrm{ Mpc}/h$ and $10 \textrm{Mpc}/h$.
Until $z=2$, the void abundance increases overall scales of voids,
since the structure formation evolves as the redshift decreases.
However, after $z=2$, the merger among small voids is important in the void evolution.
While the merger events decrease the number of small voids, they enhance the abundance of large voids.
We note that the threshold is a free parameter in our methods
although we set the threshold to the average intensity to create the mock particle distribution.  
When we increase the threshold, the total number of voids decreases. 
On the other hand, when we decrease the threshold, the abundance of voids, in particular, small-scale voids, increases
because the \texttt{VIDE} algorithm identifies the substructure in large voids due to the low threshold.

We also note that our void catalogs are obtained without any filtering.
Therefore, in our analysis, some of the small-scale voids are maybe just the Poisson noise. In such voids,
the difference of the densities between the center and the ridge of the void
is weak~(see~\citealt{Neyrinck:zobov}).
In other words, the void structure is not evolved well in such voids.
However, in our method, it is not important
whether the identified structures are true voids or not,
rather it is important whether such structures are statistically isotropic.
Therefore, we can stack the identified structure as a ``void''
without caring whether it is a true void or not.

\begin{figure*}
	\includegraphics[width=0.3\linewidth]{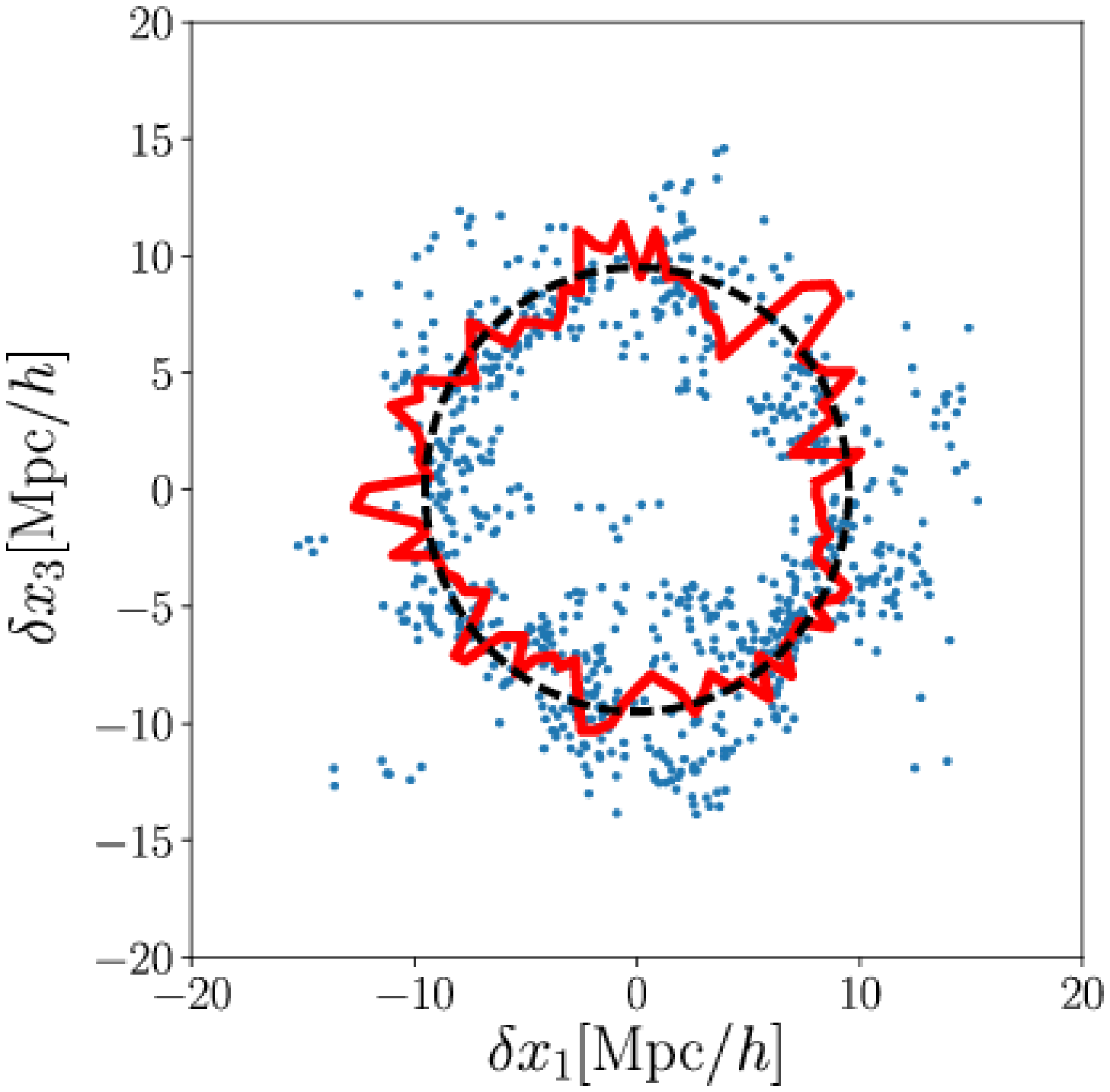}
	\includegraphics[width=0.3\linewidth]{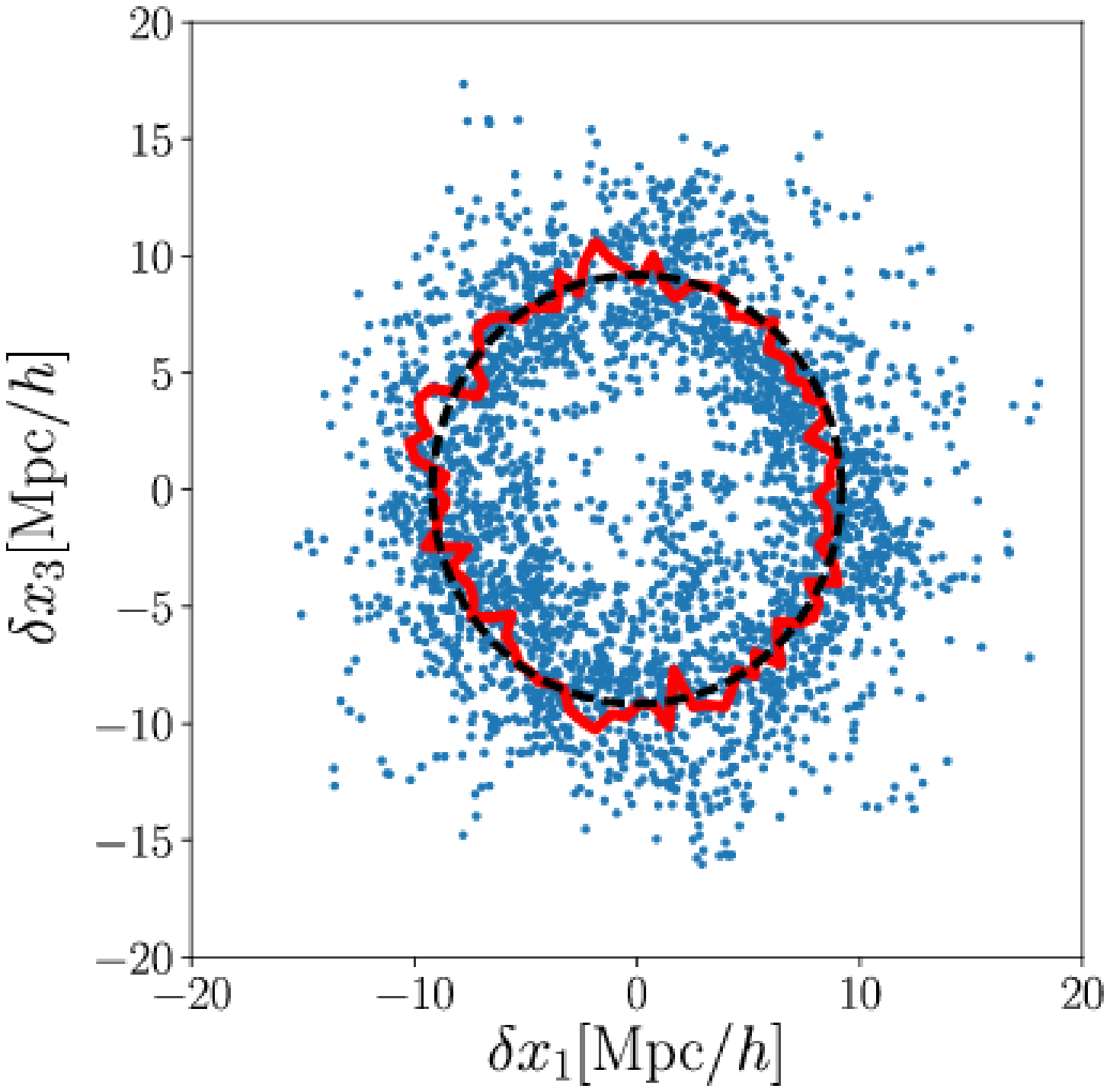}
	\includegraphics[width=0.3\linewidth]{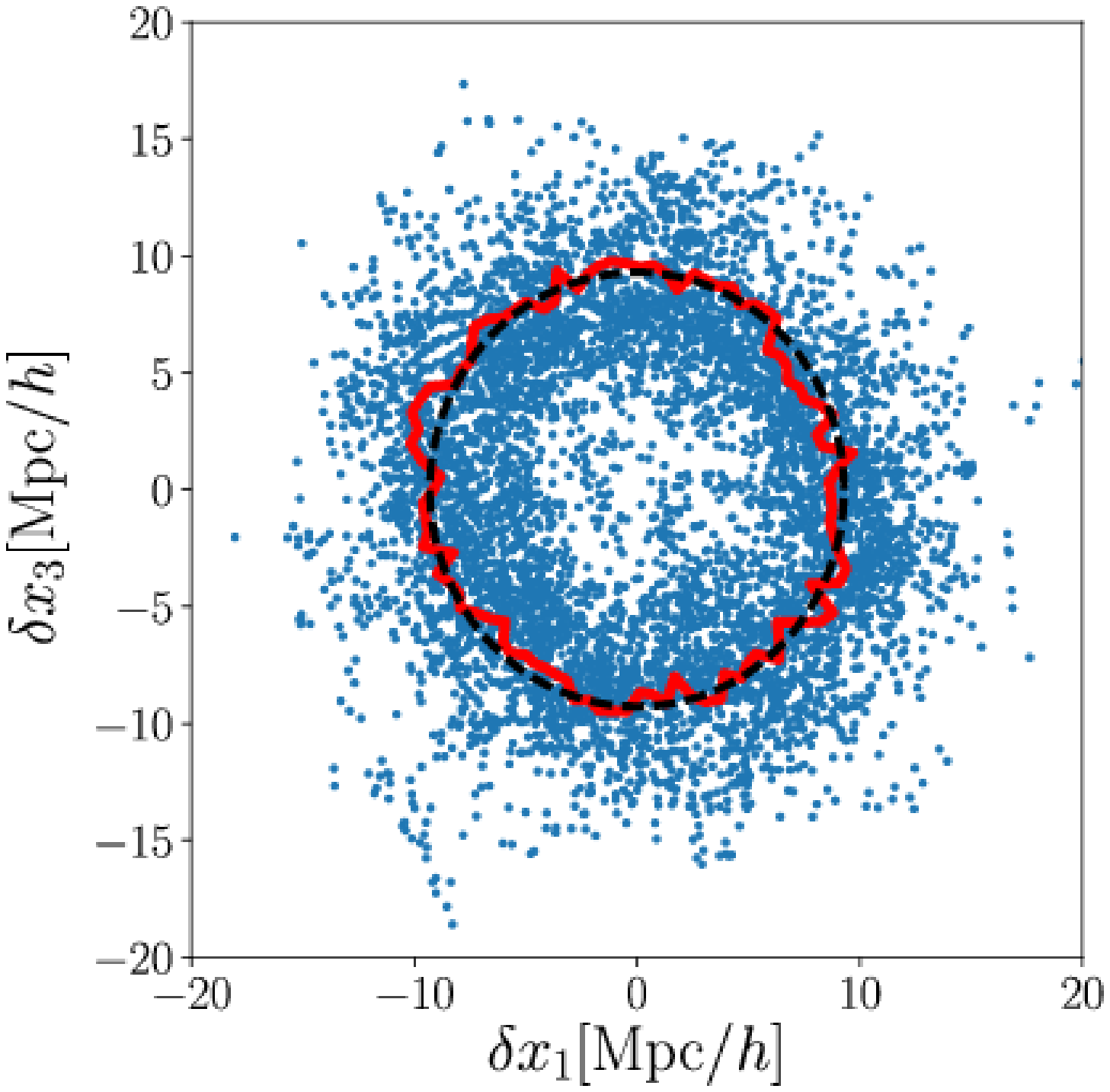}
 	\caption{
 	We show how the stacking of void works.
	From the left to the right panel, we stack $10,50$ and $100$ voids.  
	We show the 2 dimensional stacked voids with $R_{\rm eff} = 10\pm 0.5 \ [{\rm Mpc}/h]$ at $z=1$.
	The dashed black line shows
	a reference circle and the solid red line shows
	averaged particle position within an azimuthal bin. 
	The averaged profile becomes closer to the spherical shape when the number of voids is larger.
	\label{fig:stackReal}}
\end{figure*}

\begin{figure}
 	\includegraphics[width=\linewidth]{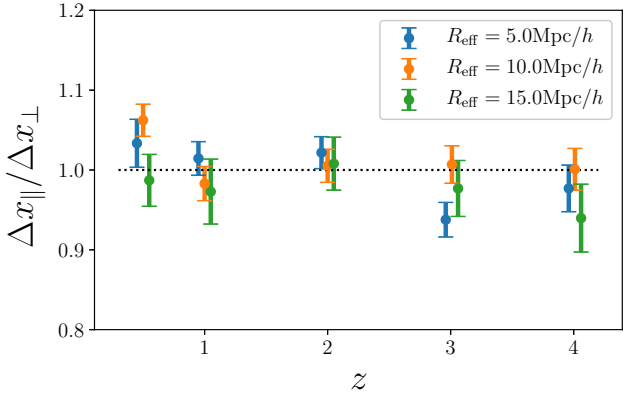}
	 \caption{ The ratio between $\Delta x_{\parallel}$ and $\Delta x_{\perp}$ of the stacked voids in the 21cm map.
	 The error bar is measured from the one realization of the simulation box.  
 One can see that the size ratio is consistent with unity, which suggests that the shape of the stacked void is spherical. 
 We also see that this property does not depend on the redshift and the size of the stacked void. 
 \label{fig:APsignal}}
\end{figure}

%%%%%%%%%%%%%%%%%%%%%%%%%%%%%%%%%%%%%%%%%%%%%%%%%%
\section{The Alcock Paczynski Test with stacked voids}
\label{sec:APtest} % used for referring to this section from elsewhere
%%%%%%%%%%%%%%%%%%%%%%%%%%%%%%%%%%%%%%%%%%%%%%%%%%
\subsection{The AP test}
\label{ssec:APtest} % used for referring to this section from elsewhere
%%%%%%%%%%%%%%%%%%%%%%%%%%%%%%%%%%%%%%%%%%%%%%%%%%
The AP test can determine the cosmological model by using the known geometrical information about the target objects~\citep{Alcock:1979}.
Suppose that we observe an object which has the comoving sizes $\Delta x_\parallel$ in the line-of-sight direction and
$\Delta x_\perp$ in the perpendicular direction.
If it is located at a cosmological distance from us, the redshift span $\Delta z$ of the object corresponding to $\Delta x_\parallel$ is
\begin{align}
	\Delta x_{\parallel} = \int^{z+\Delta z} _{z} \frac{c dz'}{H(z')}
	\approx \frac{c}{H(z)} \Delta z,
\end{align}
where $H(z)$ is the Hubble parameter,
\begin{align}
	H(z) = H_0 \sqrt{ \Omega_{{\rm m} ,0} (1+z)^3 + (1 - \Omega_{\rm m}) (1+z)^{3(1+w)}},
\end{align}
with the present energy density parameters of matter and the equation of state parameter of dark energy, $\Omega_{{\rm m}} $ and $w$, respectively.
Here we have assumed a flat universe.
We note that since we will work in the redshifts, $z \leq 4$, we can drop the contribution from the radiation components to the Hubble parameter.  
  
The observed angular scale $\Delta \theta$ of the object is given by
\begin{align}
	\Delta x_{\perp} = \chi(z) \Delta \theta,
\end{align}
where $\chi(z)$ is the comoving distance from the observer to the object at the redshift $z$,
\begin{align}
	\chi(z) = \int^z_0 \frac{c dz'}{H(z')}.
\end{align}

If the object is spherically symmetric, $\Delta x_{\parallel} = \Delta x_{\perp}$,
the observed redshift span and angular size can be related as
\begin{align}
	\frac{\Delta z}{z\Delta \theta} = \frac{1}{cz}\chi(z) H(z).
	\label{eq:APsignal}
\end{align}
Here, the left hand side is the observable,
while the right hand side can be calculated from the cosmological model.
Therefore, when the observable and the theoretical prediction
from the cosmological model satisfy the relation given in equation~\eqref{eq:APsignal},
the cosmological model used in the prediction is correct to describe our Universe.

%%%%%%%%%%%%%%%%%%%%%%%%%%%%%%%%%%%%%%%%%%%%%%%%%%
\subsection{Void stacking}
\label{ssec:voidStack} % used for referring to this section from elsewhere
%%%%%%%%%%%%%%%%%%%%%%%%%%%%%%%%%%%%%%%%%%%%%%%%%%
According to the cosmological principle, we can make the hypothesis 
that the voids found in the 21cm intensity map as described in the previous section are also statistically spherical symmetric.
To check the hypothesis, we stack voids found in the 21cm map and evaluate the size ratio between
$\Delta x_{\parallel}$ and $\Delta x_{\perp}$ of the stacked void.
In the stacking process, we convert the particle positions in the simulation box
into the relative position from the center of the void, 
\begin{align}
	\delta \bm{x}_i = \bm{x}_i - \bm{x}_c.
\end{align}
where $\bm{x}_i$ and $\bm{x}_c$ are the positions of the $i$-th particle and the center of the void to which the $i$-th particle belongs.  
The position of the center of a void is determined as the Voronoi cell volume-weighted center,
\begin{align}
	\bm{x}_c = \frac{\sum V_i\bm{x}_i}{\sum V_j}, 
\end{align}
where $V_i$ represents the Voronoi cell volume of the $i$-th particle.  

Then, we take the second moment of the stacked particle distribution to determine the size of the stacked void.  
We assume that the $x_3$ axis is the line-of-sight direction, and $x_1$ and $x_2$ are the perpendicular direction.
In this configuration, the second moments for each direction are evaluated as
\begin{align}
	\Delta x_{\parallel}^2 &= \braket{\delta x_{3}^2 }=\frac{1}{N} \sum_i^N \delta x_{i,3}^2,
	\label{eq:parallel_x}\\
	\Delta x_{\perp}^2 &= \frac{\braket{\delta x_{1}^2} + \braket{\delta x_2^2}}{2}=\frac{1}{N} \sum_i^N \frac{(\delta x_{i,1}^2 + \delta x_{i,2}^2 )}{2} ,\label{eq:perpendicular_x}
\end{align}
where $N$ is the total number of particles to be stacked.  
We stack voids only within the narrow radial size of 
$\Delta R_{\rm eff}=1 {\rm Mpc}/h$, where $5\leq R_{\rm eff}\leq15 {\rm Mpc}/h$.
The information of different size voids is merged in the level of the likelihood as discussed in section \ref{ssec:parameterEstimation}.

We first visualize the shape convergence according to the number of voids in the 2-dimensional stacking in Figure~\ref{fig:stackReal}.  
Here we show the stacked void of the radius $R_{\rm eff} = 10 {\rm Mpc}/h$ at $z=1$.  
We increase the number of voids for stacking as $10$, $50$ and $100$ from the left panel to the right panel respectively.
The red solid lines show the averaged particle positions in each sector while the black dotted lines indicate the reference circles.  
One can see that the shape of the stacked void is noisy when the number of voids is small.
On the other hand, the averaged void shape is likely to be spherical when the number of voids is large.  

Then, we plot the ratio between $\Delta x_{\parallel}$ and $\Delta x_{\perp}$ in Figure \ref{fig:APsignal}.
In Figure \ref{fig:APsignal} we represent the results of the stacked voids of radii $R_{\rm eff}=5,10$ and $15{\rm Mpc}/h$.  
To evaluate the error bars, we measure the variance by 100 times bootstrapping, which was sufficient to converge.

The results show us that the statistical error depends on the number of voids. 
Therefore, the error bar becomes large as the number of voids decreases as shown in Figure \ref{fig:numberFunction}.
However, the ratio does not deviate from unity more than a $2$-$\sigma$ confidence level.
This fact strongly supports our hypothesis that the void shape
in the contour maps is statistically spherical.
Therefore, the voids in the 21cm contour maps are appropriate for the AP test.

We note that the future SKA intensity mapping survey such as SKA1-MID will cover 
$20,000 [\textrm{deg}^2]$ in the sky \citep{SKA:redbook2018}.  
This sky coverage corresponds to 20 times larger than the simulation size at $z = 0.5$.  
For higher redshifts, the comoving volume also becomes larger. 
Thus the statistical error may be roughly reduced by at least $1/\sqrt{20}$ times for the future survey.
Therefore, in future surveys, the actual statistical uncertainty for the shape ratio
could be smaller than our estimation in Figure~\ref{fig:APsignal}.

%%%%%%%%%%%%%%%%%%%%%%%%%%%%%%%%%%%%%%%%%%%%%%%%%%
\subsection{Parameter estimation}
\label{ssec:parameterEstimation} % used for referring to this section from elsewhere
%%%%%%%%%%%%%%%%%%%%%%%%%%%%%%%%%%%%%%%%%%%%%%%%%%
\begin{figure}
	\includegraphics[width=1.0\linewidth]{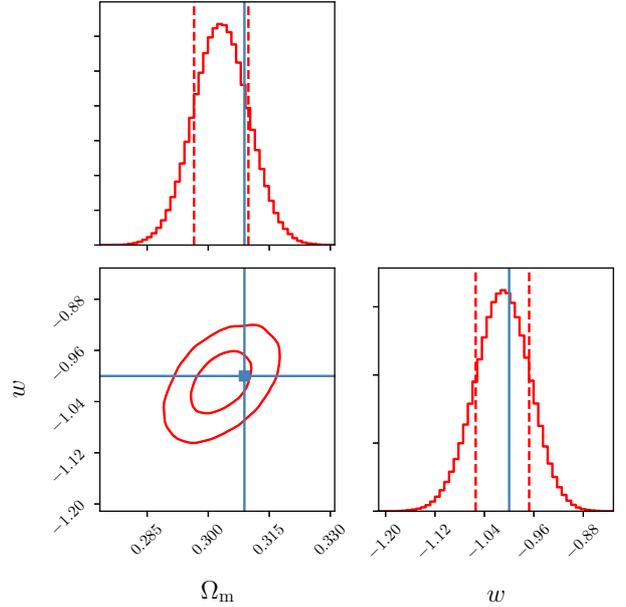}
 	\caption{
 	Expected parameter constraints given our simulation volume in the real space. The expected error is obtained by stacking all voids found in the simulation volume. 
 	For 2-D contour, the circles correspond 1 and 2-$\sigma$ confidence regions and 1-D histogram is a likelihood marginalized over other parameters. The best-fitting value for each parameter is
 	$\Omega_{\rm m} = 0.3031^{+0.0067}_{-0.0066}$ and $w=-1.010^{+0.043}_{-0.044}$.
 	 Our fiducial value is indicated by the blue dot. 
 	As we see in Figure~\ref{fig:APsignal}
 	%,fig:shapeConvergenceReal}, 
 	the stacked void is well described by
 	   the sphere so that the AP test correctly measures the cosmological parameters.
	\label{fig:MCMCReal}}
\end{figure}

\begin{figure*}
	\includegraphics[width=0.3\linewidth]{stack_N100_realSpace.eps}
	\hspace{10mm}
	\includegraphics[width=0.3\linewidth]{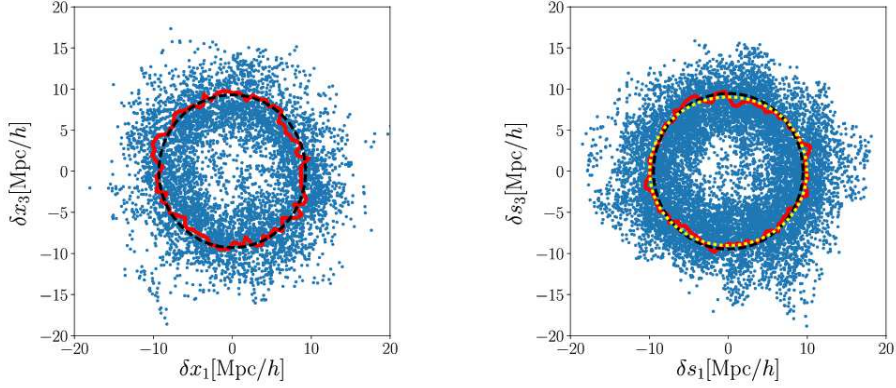}
	\caption{The comparison of stacked void shape in real-space and redshift-space.  
	The left panel is the shape of the voids in real-space and the right panel shows the one in redshift-space, after stacking over 100 voids.
 	The yellow dotted line indicates the ellipsoid which is fitted to the red line, where the best-fitting ratio between the major and minor axis being 0.9075.
	\label{fig:stackRedshift}}
\end{figure*}

 \begin{figure}
	 \includegraphics[width=\linewidth]{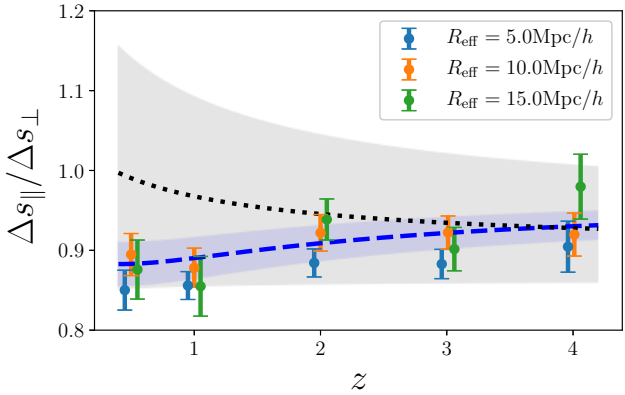}
 	 \caption{
	 The same as Figure \ref{fig:APsignal}, but in redshift-space.
	 The redshift-space distortion effect does not depend on the void size but redshift.
	 The blue dashed line is 
	 the best-fitting curve of equation \eqref{eq:offset} with keeping cosmological parameters to their input values.
	 The black dotted line is the one with all parameters including $\Omega_{\rm m}$ and $w$ are simultaneously fitted.
	 \label{fig:APsignal-redshift}}
 \end{figure}

Let us demonstrate the AP test of the stacked voids in the 21cm contour maps to determine the cosmological parameter.
Here we adopt the Markov Chain Monte Carlo method which is a powerful tool to estimate the parameter from a set of observation signals.
This method provides us with the posterior probability of a model parameter based on the Bayes' theorem when we obtain a data set by observations.  
Suppose we conduct an observation and obtain the observation data set $\bm e$, 
the posterior probability of the parameter $\bm p$ is expressed by the multiplication of the likelihood and the prior probability of the parameter, 
\begin{align}
	P'(\bm{p} | \bm{e}) \propto L(\bm{p} | \bm{e}) \times P(\bm{p}),
\end{align}
where $P'(\bm{p} | \bm{e})$ is the posterior probability, $L(\bm{p} | \bm{e})$ is the likelihood,
and $P(\bm{p})$ is the prior probability of the parameter.  
If we assume that the distribution of the observation data follows the Gaussian distribution, the likelihood from these measurements is given by
\begin{align}
	%L = \prod_{i,j} \exp \left[ - \frac{\left( (\Delta z / z \Delta \theta)|^{\textrm{data}}_{i,j} - (\Delta z / z \Delta \theta)|^{\textrm{Theory}}_{i,j} \right)^2}{2 \sigma^2_{i,j}} \right].
	L (\bm{p}|\bm{e})
	=
	 \prod_{i,j} \frac{1}{\sqrt{2\pi \sigma^2_{i,j}} } \exp \left[ -\frac{ \left( e^{\textrm{data}}_{i,j} - e_i(\bm{p}) \right)^2 }{2 \sigma^2_{i,j}} \right],
	\label{likelihood}
\end{align}
where $e^{\textrm{data}}_{i,j}$ and $\sigma^2_{i,j}$ are
the measured value and the variance of the observation data of the stacked void of the radius $ R_{{\rm eff},j}$ at $z_i$
and $e_i(\bm{p})$ is the theoretical predicted value for the component with the parameter $\bm p$ at $z_i$.
In this demonstration of the AP test, $z_i$ means one of the redshift bins while 
$R_{{\rm eff},j}$ means one of the radii of the stacked void, where $R_{\rm eff} = 5,6,\cdots,15 {\rm Mpc}/h$.
$\bm{p}$ includes a set of $\Omega_{\rm m}$ and $w$. Hence the theoretical prediction $e_i(\bm{p})$ is evaluated as
\begin{align}
	e_i(\Omega_{\rm m}, w) \equiv	\left.
	\frac{\Delta z}{z\Delta \theta} \right|_{{\rm theory},i}= \frac{1}{cz_i}\chi(z_i,\Omega_{\rm m},w) H(z_i,\Omega_{\rm m},w).  
\end{align}
On the other hand, $e^{\textrm{data}}_{i}$ is obtained through
\begin{align}
	e^{\textrm{data}}_{i,j} \equiv	\left. \frac{\Delta z}{z\Delta \theta} \right|_{{\rm data},i,j} = \frac{1}{cz_i}\chi(z_i) H(z_i)
 \sqrt{\frac{\Delta x_{\parallel}^2(z_i,R_{{\rm eff},j})}{\Delta x_{\perp}^2(z_i,R_{{\rm eff},j})} },
\end{align}
where we use the mean value in Figure~\ref{fig:APsignal} as $\Delta x_{\parallel}^2/{\Delta x_{\perp}^2}$
and the fiducial values of matter density and equation of state parameter,
$\Omega_{\rm m,fid} =0.3089$ and $w_{\rm fid}=-1$,
which are adopt in the IllustrisTNG300-3.
For the MCMC analysis, we use a python module \texttt{EMCEE} \citep{EMCEE}.

We plot the probability distribution of the parameters obtained by the MCMC analysis in Figure \ref{fig:MCMCReal}.  
The results show that the parameter estimation is consistent with the fiducial values within a 1-$\sigma$ confidence level.  
The estimated values with 1-$\sigma$ errors are $\Omega_{\rm m} = 0.3031^{+ 0.0067}_{-0.0066}$
and $w=-1.010^{+0.043}_{-0.044}$.
Especially, the estimation of $w$ is very close to the fiducial value while that of $\Omega_{\rm m}$ is barely consistent with the 1-$\sigma$ region.  
The estimation of matter density has uncertainty about 2\%.  
This result is derived by only the AP signals of the stacked void.  
The current joint analysis of the CMB power spectrum, CMB lensing, and BAO reported
the uncertainty on $\Omega_{\textrm{m}}$ is about $2\%$ \citep{Planck2018:cosmology}
when they assume the $\Lambda$CDM model.
our analysis result achieves almost the same level estimation as the current detailed cosmological observation.  
Therefore, the AP test with stacked voids in the 21cm intensity maps has the potential to provide stronger constraints
on the cosmological parameters rather than AP test with stacked voids in galaxy distributions.

Before ending this section, we comment on the observation noise.
In this work, we do not consider any contamination such as foreground effects or equipment noises,
and the angular and frequency resolution of radio telescopes.
In our method, we use the contour line corresponds to the threshold to trace the shape of the voids.  
Although we set the threshold to the mean value in the map,
we can freely choose this threshold.
To identify the shapes of the structures in actual observations,
it is required that the threshold is larger than the noise level.
In lower redshifts,~$z<5$, the dominant noise contribution
comes from the thermal noise of observation equipment. 
We confirm that the noise level might be higher than the means signals
in this observation set up at $1<z$ when we assume the instrumental noise~\citep{Horii:2017}.  
In this case, we can set the threshold to the higher value so that we can trace the void structures.  
 \begin{figure*}
	 \includegraphics[width=0.8\linewidth]{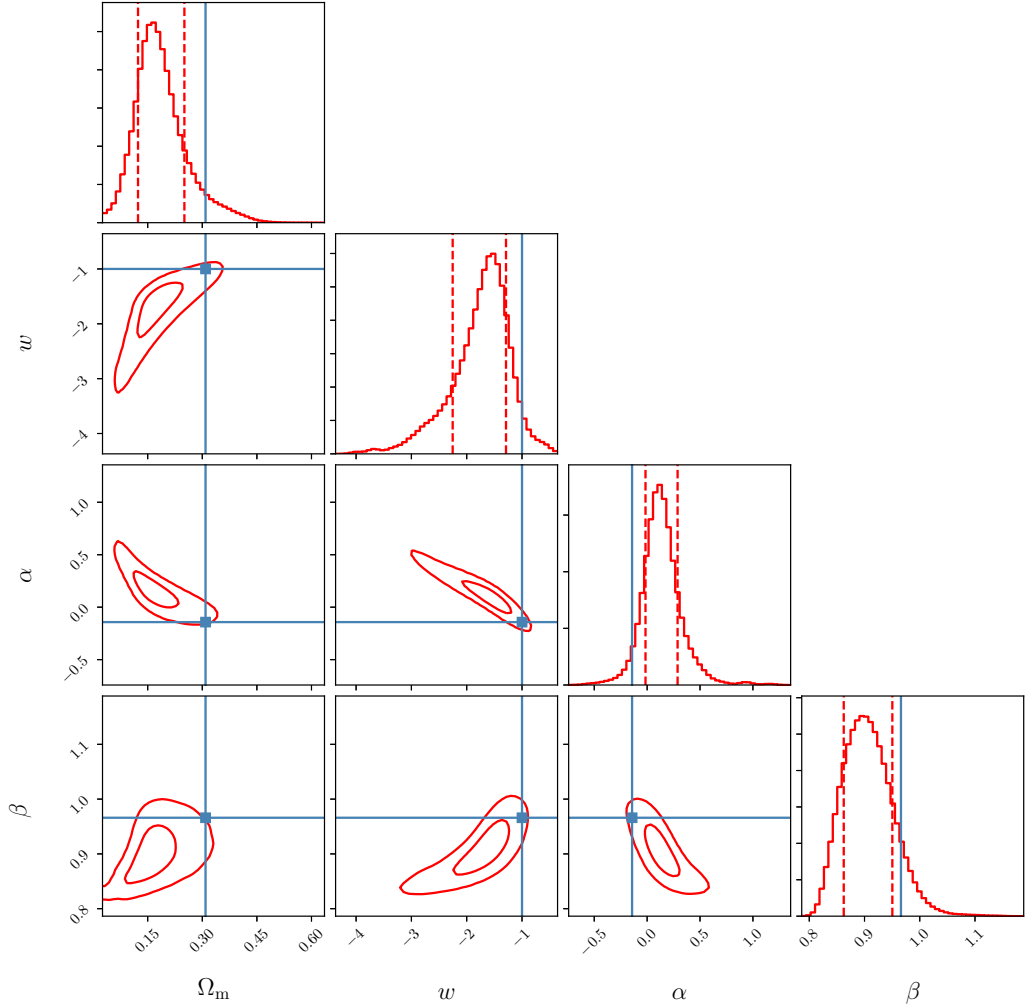}
 	 \caption{
 	 Same as Figure \ref{fig:MCMCredshift} but all the parameters are jointly fitted simultaneously. We find that the RSD correction parameters and cosmological parameters are largely degenerate with each other. The estimated value for the cosmological parameters are biased; however, it is still consistent within 2$\sigma$.
 	 The best-fitting values are
	 $\Omega_{\rm m} = 0.1723^{ + 0.0734}_{ - 0.0560}$,
	 $w=-1.692^{ + 0.385}_{ - 0.663}$,
	 $\alpha = 0.1329^{ + 0.1830}_{ - 0.1409}$, and 
	 $\beta = 0.9016^{ + 0.0480}_{ - 0.0427}$.	  
	 \label{fig:MCMC_redshift_full}}
 \end{figure*}

%%%%%%%%%%%%%%%%%%%%%%%%%%%%%%%%%%%%%%%%%
\section{redshift space distortion effect}
%%%%%%%%%%%%%%%%%%%%%%%%%%%%%%%%%%%%%%%%%
The peculiar velocity of the $\HI$ gas gives two effects on the 21cm signals.
One is the broadening of the line spectrum, which contributes through $V_{\parallel}$ in equation~\eqref{eq:delta_Tb}.
As mentioned above, this effect is subdominant compared with
the Hubble velocity. Therefore, we neglect it in this paper.

The other effect is the redshift-space distortion~(RSD).
The observed positions of the intensity signals are modified along
the line-of-sight direction because the redshift is affected by Hubble expansion as well as Doppler shift due to the peculiar velocity, which cannot be discriminated.  
In this section, we discuss the RSD effect on the void findings and the AP test in the 21cm intensity maps.

To take into account the RSD effect, we construct the intensity maps in redshift-space.
The procedures are the same as in the real-space but first of all, we shift the position of gas particles according to their peculiar velocities along the line-of-sight direction,
\begin{align}
	s_\parallel = x_\parallel + \frac{(1+z)\upsilon_{{\rm p}\parallel}}{H(z)}.
	\label{eq:redshift_space}
\end{align}
Then gridding, converting to the brightness temperature, particle re-distribution and void finding procedures are all same as in the real-space.

%%%%%%%%%%%%%%%%%%%%%%%%%%%%%%%%%%%%%%%%%
\subsection{Void shape in redshift-space}
\label{ssec:Shape_redshiftSpace}
%%%%%%%%%%%%%%%%%%%%%%%%%%%%%%%%%%%%%%%%%

In redshift-space, we have confirmed that the stacked voids become squashed along the line-of-sight, 
which is also found in the previous studies in dark matter density fields or mock galaxy distributions
\citep{Lavaux:2012,Mao:2017}.
Figure \ref{fig:stackRedshift} demonstrates how the stacked void shape is deformed in the redshift-space (right) compared to the one in the real-space (left).
For both panels, we set vertical axes as the line-of-sight.  
The yellow dotted curve in the right panel shows the best-fitting ellipse to the measured profile shown by the red solid line.
Compared to the perfect circle shown by the black dashed line, the ellipse is squashed along the line-of-sight direction.

To show the impact of the RSD effect on the shape deformation, we plot the ratio, $\Delta s_{\parallel}/\Delta s_{\perp}$,
in Figure~\ref{fig:APsignal-redshift}.
The RSD effect appears by about 10~\%-level flattening of stacked voids along the line-of-sight direction.  
It does not depend on the void size but does on the redshift.
The 10\% distortion due to the RSD has been already found in the previous works
with N-body simulations, where dark matter particles or mock galaxies are used as a tracer of the void \citep{Lavaux:2012,Mao:2017}.
The previous works reported that the shape distortion is about 10 to 15\% 
at $z<1$, which is consistent with our results.

%%%%%%%%%%%%%%%%%%%%%%%%%%%%%%%%%%%%%%%%%
\subsection{AP test in redshift-space}%with the RSD effect}
\label{ssec:APtest_redshiftSpace}
%%%%%%%%%%%%%%%%%%%%%%%%%%%%%%%%%%%%%%%%%

Here we present the cosmological parameter estimation 
using the AP test in the presence of the RSD effect.
If we do not consider the RSD effect in redshift-space,
$\Omega_{\rm m}$ and $w$ are both underestimated, since
smaller values of $\Omega_{\rm m}$ and $w$ make AP signal 
(equation \eqref{eq:APsignal} ) smaller.
Previous works attempt to correct the RSD effect by multiplying 
a constant factor \citep{Lavaux:2012, Sutter:2014, Mao:2017}
and make the parameter estimation unbiased.
However, as can be seen in Figure \ref{fig:APsignal-redshift}, 
the offset from unity is not constant in our case simply 
because we consider a wide range of redshift at the same time.
We find that the factor of $0.9071^{+0.0032}_{-0.0031}$ is reasonable to describe the offset so that 
we need to divide the AP signals by this factor to correct for the RSD effect with the constant offset.
With this correction, the estimated cosmological parameters 
are highly biased, which reflects the model of correcting 
the RSD effect by the constant factor is not appropriate.

In our analysis, we assume that the deformation depends on the redshift
due to the velocity evolution.
To model the time dependence, we expand $\Delta s_{\parallel}/\Delta s_{\perp}$ as
\begin{align}
	\frac{\Delta s_{\parallel}^2}{\Delta s_{\perp}^2} 
	=
	 \frac{1}{\Delta x_{\perp}^2} 
	 \left[ \braket{\delta x_{\parallel}^2} + 2\frac{(1+z) \braket{\delta x_{\parallel} \cdot \upsilon_{{\rm p} \parallel}}}{H(z)} + \frac{(1+z)^2 \braket{\upsilon_{{\rm p} \parallel}^2}}{H(z)^2} \right].  
\end{align}
According to the linear theory, the peculiar velocity is related to the density perturbation,
\begin{align}
	\bm{\upsilon}_{\rm p}(\bm{x}) = \frac{H(z) f(z) D(z)}{(1+z)}\int d^3 x' \frac{\bm{x} - \bm{x}'}{|\bm{x} - \bm{x}'|^3}
	\delta(\bm{x}'),
\end{align}
where $f(z)$ and $D(z)$ are the linear growth rate and the linear growth factor, respectively. 
With some calculation, we find that $\Delta s_{\parallel}/\Delta s_{\perp}$ can be formally written by
\begin{align}
	\frac{\Delta s_{\parallel}}{\Delta s _{\perp}}
	=
	\alpha f(z)D(z) + \beta,
	\label{eq:offset}
\end{align}
with constants $\alpha$ and $\beta$. This can be used to correct for the 
RSD effect instead of the constant offset. In the case where the perturbation is well within the linear regime and in the absence of velocity biases, the parameters $\alpha$ and $\beta$ can be predictable from the linear theory. However, in practice, the values predicted by the linear theory do not explain the amount of deformation by the RSD and we leave them as free parameters.

First, we fit the AP signal with cosmological parameters and calibration parameters in equation \eqref{eq:offset} simultaneously.
In Figure \ref{fig:MCMC_redshift_full}, we show the constraints on those parameters. 
The blue solid lines represent the fiducial values. The fiducial values for nuisance parameters $\alpha$ 
and $\beta$ can be defined later in this section.
One can see that the estimations deviate from the fiducial values when we search the preferable parameters at the same time, in more than 1-$\sigma$ but less than 2-$\sigma$.
In this case, $\Omega_{\rm m}$ and $w$ are under estimated by the parameter search such that 
$\Omega_{\rm m} = 0.1723^{ + 0.0734}_{- 0.0560}$ and $w = -1.692^{+0.385}_{-0.663}$.
The reason for these wrong estimations comes from the incorrect estimations of the correction parameters, $\alpha$ and $\beta$.  
According to the results, we obtain $\alpha = 0.1329^{ + 0.1830}_{ - 0.1409}$ and $\beta=0.9016^{ + 0.0480}_{ - 0.0427}$. 
In Figure \ref{fig:APsignal-redshift}, we plot the offset function as the black dotted line.  
One can see that the offset function does not well trace the data, particularly at low redshifts.  
Therefore, the estimated cosmological parameters are biased. 
As shown in Figure \ref{fig:MCMC_redshift_full}, the nuisance parameters degenerate with the cosmological parameters, which indicates that the tight and correct constraints on the calibration are required to make the estimate accurate and precise.

Then, we would like to see whether this calibration model may work when the parameters are properly provided a priori. To do this, we fix the cosmological parameters to their fiducial values and find the best-fitting values for $\alpha$ and $\beta$, which results in $\alpha= -0.1424^{-0.0271}_{-0.0272}$ and $\beta=0.9660^{+0.0118}_{-0.1160}$. We refer to these values for $\alpha$ and $\beta$ as fiducial values aforementioned. The predicted distortion is shown in Figure \ref{fig:APsignal-redshift} with blue dashed line. The prediction nicely recovers the observed distortion at all redshift ranges. Given the fiducial values for $\alpha$ and $\beta$, we run the likelihood analysis only for the cosmological parameters. 
Then we have 
the best-fitting parameters as $\Omega_{\rm m} = 0.3093^{+0.0076}_{-0.0077}$ and $w=-1.008^{+0.042}_{-0.044}$, 
which are fairly consistent with the fiducial values by the same uncertainty level as in the case without the RSD effect. Therefore, as long as we can calibrate the RSD effect via simulation as prior knowledge, the model well describes the RSD correction and we can measure the cosmological parameters in an unbiased manner. 

\begin{figure}
	\includegraphics[width=1.0\linewidth]{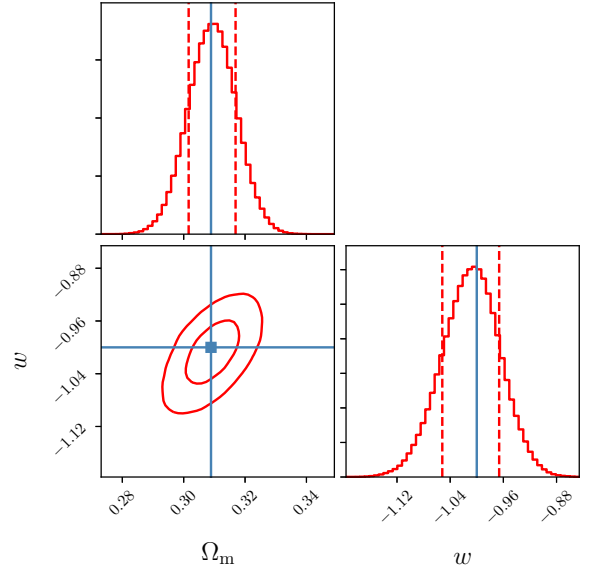}
 	\caption{
 	Expected errors at redshift-space. We correct for the redshift-space distortion by equation \eqref{eq:offset}. The values for $\alpha$ and $\beta$ are obtained by assuming $\Omega_{\rm m}$ and $w$ being at their fiducial values. After the correction, we estimate the cosmological parameters and find their best-fitting values as  
 	$\Omega_{\rm m} = 0.3093^{+0.0076}_{-0.0077}$ and $w=-1.008^{+0.042}_{-0.044}$.
 \label{fig:MCMCredshift}}
\end{figure}

%%%%%%%%%%%%%%%%%%%%%%%%%%%%%%%%%%%%%%%%%%%%%%%%%%
%%%%%%%%%%%%%%%%%%%%%%%%%%%%%%%%%%%%%%%%%%%%%%%%%%
\section{Conclusion}
In this paper, we have presented the AP test with the 21cm voids
as a new cosmological parameter estimation
in future 21cm observations.
To demonstrate the AP test,
we have constructed the 21cm intensity maps by using the state-of-the-art cosmological hydrodynamics simulation,
the IllustrisTNG300-3. Then, using the void finder algorithm \texttt{VIDE}, we made void catalogs in which the void shapes are identified through tracing the critical intensity contour.
The shape of the individual void is far from spherical.  
However, we have shown that the stacked void becomes more spherical as increasing the number of voids to be stacked.
This result suggests that the voids in the 21cm intensity maps are appropriate for the AP test.

To present the potential of the AP test with 
the stacked void of the 21cm intensity maps,
we have performed 
the Markov Chain Monte Carlo analysis for
the parameter estimation on the matter-energy density parameter and the equation of state of dark energy.
Our result shows that the parameter estimation by the AP test with 21cm stacked voids
is consistent with the fiducial values within 1-$\sigma$
confidence level.  
In particular, in the estimation of the matter-energy density parameter,
the uncertainty level can be controlled in about 2\%,
which is the same level as the result of the joint analysis among CMB temperature anisotropy, CMB lensing, and BAO.

Similar to the case with voids in galaxy maps, 
the RSD effect is one of the challenges in the AP test with 21cm intensity maps. 
The peculiar velocities of neutral hydrogen gas deform the void shapes in 21cm intensity maps.
To investigate the impact of the RSD effect on the AP test, we construct
the redshift-space 21cm intensity maps with the RSD effect.
We found that the RSD effect arises as 
the flattening effect on the void shape along the line-of-sight direction
by about 10\% level, compared to the intensity map without the RSD effect.
To remove the RSD effect in the AP test,
we have suggested the correction factor whose redshift evolution is motivated from the one of the peculiar velocity in the linear perturbation theory.
We have shown that we can correct the RSD effect and recover the same uncertainty level as in the case without the RSD effect when we successfully calibrate the correction factor.  
Thus, our results strongly encourage to apply stacked voids in the $\HI$ intensity maps to the AP test.  
We will conduct further investigation of how we calibrate the correction factor by performing the numerical simulations with different cosmological models.

%%%%%%%%%%%%%%%%%%%%%%%%%%%%%%%%%%%%%%%%%%%%%%%%%%
\section*{Acknowledgements}
%%%%%%%%%%%%%%%%%%%%%%%%%%%%%%%%%%%%%%%%%%%%%%%%%%
We would like to thank the supports of MEXT's Program for Leading Graduate Schools Ph.D. professional,
``Gateway to Success in Frontier Asia".
This work is supported by MEXT KAKENHI Grant Number 15H05890.
%The Acknowledgements section is not numbered. Here you can thank helpful
%colleagues, acknowledge funding agencies, telescopes, and facilities used, etc.
%Try to keep it short.

%%%%%%%%%%%%%%%%%%%%%%%%%%%%%%%%%%%%%%%%%%%%%%%%%%

%%%%%%%%%%%%%%%%%%%% REFERENCES %%%%%%%%%%%%%%%%%%

% The best way to enter references is to use BibTeX:

%\bibliographystyle{mnras}
%\bibliography{example} % if your bibtex file is called example.bib

% Alternatively you could enter them by hand, like this:
% This method is tedious and prone to error if you have lots of references
%\begin{thebibliography}{99}
%\bibitem[\protect\citeauthoryear{Author}{2012}]{Author2012}
%Author A.~N., 2013, Journal of Improbable Astronomy, 1, 1
%\bibitem[\protect\citeauthoryear{Others}{2013}]{Others2013}
%Others S., 2012, Journal of Interesting Stuff, 17, 198
%\end{thebibliography}

\bibliographystyle{mnras}
\bibliography{bibdata}

%%%%%%%%%%%%%%%%%%%%%%%%%%%%%%%%%%%%%%%%%%%%%%%%%%

% Don't change these lines
\bsp	% typesetting comment
\label{lastpage}
\end{document}